\documentclass[pteplogo]{ptephy_v2} 

\usepackage{graphicx}
\usepackage{url} 

\usepackage{booktabs}
\usepackage{lastpage}

\usepackage{amsmath}
\usepackage{amsfonts}
\usepackage{amssymb}
\usepackage{amsthm}
\usepackage{amstext}
\usepackage{array}
\usepackage{bm}
\usepackage{dcolumn}
\usepackage{enumerate}
\usepackage{lineno}
\usepackage{mathtools}
\usepackage{multirow}
\usepackage{orcidlink}
\usepackage[normalem]{ulem}
\usepackage{scalerel}
\usepackage{textcomp}
\usepackage{txfonts}
\usepackage{ulem}
\usepackage{xcolor}

\usepackage{anyfontsize}

\allowdisplaybreaks

\usepackage{hyperref}
\hypersetup{
    colorlinks=true,
    linktocpage,
    unicode,
    linkcolor=blue,
    filecolor=magenta,      
    urlcolor=cyan
}

\definecolor{CiteColor}{rgb}{0,0.5,0}
\hypersetup{citecolor=CiteColor}

\definecolor{red  }{rgb}{1,0,0}
\definecolor{blue }{rgb}{0,0,1}
\definecolor{green}{rgb}{0,1,0}
\definecolor{magenta}{rgb}{1,0,1}
\definecolor{rfgreen}{RGB}{102,170,85}

\nolinenumbers

\begin{document}
\title{
Secular evolution of orbital parameters for general bound orbits in Kerr spacetime
}

\author[1]{Norichika~Sago\,\orcidlink{0000-0002-1430-5652}}
\author[2,3]{Ryuichi~Fujita\,\orcidlink{0000-0002-8415-8168}}
\author[4]{Soichiro~Isoyama\,\orcidlink{0000-0001-6247-2642}}
\author[5]{Hiroyuki~Nakano\,\orcidlink{0000-0001-7665-0796}}

\affil[1]{Division of General Education, Kanazawa Medical University, Kanazawa 920-0293, Japan}
\affil[2]{Institute of General Education, Otemon Gakuin University, Osaka 567-8620, Japan}
\affil[3]{Center for Gravitational Physics and Quantum Information, Yukawa Institute 
for Theoretical Physics, Kyoto University, Kyoto 606-8502, Japan}
\affil[4]{Department of Physics, National University of Singapore, Singapore 117551, Singapore}
\affil[5]{Faculty of Law, Ryukoku University, Kyoto 612-8577, Japan}

\begin{abstract}

We analytically derive the secular changes of the orbital parameters, i.e., energy, angular momentum, and Carter constant, for general bound orbits in Kerr spacetime, at leading order in the mass ratio, through the 6th post-Newtonian (6PN) order and the 16th order in orbital eccentricity. We validate the formulas against high-precision numerical Teukolsky results and quantify how eccentricity affects both the achievable accuracy and the PN convergence. We then construct and test a simple ``hybrid'' approximation that combines different PN and eccentricity truncations to retain accuracy at reduced computational cost. We also assess the performance of exponential resummation at higher PN orders. These results provide building blocks for fast, (analytic) adiabatic inspiral and waveform models for extreme mass ratio inspirals relevant to space-based detectors such as the Laser Interferometer Space Antenna (LISA).

\end{abstract}


\subjectindex{E01, E02, E31, E36, F31}

\maketitle


\section{Introduction}

In gravitational-wave (GW) astronomy, modelling the two-body problem in general relativity remains a central theoretical challenge. Future space-based observatories, such as LISA~\cite{LISA:2017pwj,LISA:2024hlh}, Taiji, and TianQin~\cite{2021PTEP.2021eA108L,TianQin:2020hid,Gong:2021gvw}, will probe the millihertz frequency band, targeting extreme mass ratio inspirals (EMRIs) as primary sources~\cite{Amaro-Seoane:2012lgq,LISA:2022yao,Li:2024rnk}. Concurrently, black hole perturbation theory (BHPT) and gravitational self-force (GSF) formalism~\cite{Mino:1997bw,Poisson:2011nh,Harte:2014wya,Barack:2018yvs,Pound:2021qin}---a systematic expansion of Einstein's field equations in the small mass ratio of the binary---have proven increasingly valuable for modelling intermediate mass ratio inspirals (IMRIs), which are key targets for proposed decihertz missions like DECIGO~\cite{Kawamura:2020pcg} and third-generation ground-based detectors (e.g., Einstein Telescope~\cite{ET:2025xjr} and Cosmic Explorer~\cite{Evans:2021gyd}). 
Because highly asymmetric-mass systems such as IMRIs and EMRIs complete $\sim 10^3$--$10^6$ orbital cycles during an observation period (typically ranging from a few months to several years), theoretical waveform templates must maintain sub-radian phase accuracy to prevent severe biases in parameter estimation~\cite{Burke:2023lno,Khalvati:2024tzz} and to mitigate potential losses in detection efficiency during the search phase~\cite{Mathews2025eccentric}.

To meet these stringent requirements, the LISA Waveform Working Group and the broader community are actively pursuing waveform models based on this GSF framework. 
By employing a multiscale expansion to capture the system's long-term evolution~\cite{Hinderer:2008dm,Miller:2020bft,Pound:2021qin,Mathews:2025txc,Lewis:2025ydo}, recent advances have enabled the calculation of leading adiabatic waveforms~\cite{Hughes:2021exa,Isoyama:2021jjd,Kerachian:2023oiw,Gliorio:2026yvh}, (first-order) conservative GSF effects~\cite{vandeMeent:2017bcc,Nasipak:2025tby}, and secondary-spin effects~\cite{Skoupy:2023lih,Drummond:2023wqc,Drummond:2026haw,Skoupy:2026ewu,Cui:2026qsk,Kakehi:2026up} for generic (eccentric and precessing) orbits in Kerr spacetime. Furthermore, the first post-adiabatic (1PA) waveforms, incorporating second-order GSF corrections~\cite{Pound:2019lzj,Warburton:2021kwk} with precessing secondary spins~\cite{Mathews:2025txc}, have now been achieved for quasi-circular asymmetric-mass inspirals~\cite{Wardell:2021fyy,Honet:2025gge,Honet:2025lmk} (see also Refs.~\cite{Albertini:2022dmc,Albertini:2022rfe,Albertini:2023aol,Albertini:2024rrs,Wittek:2024gxn,QiuXin:2024kup,Rink:2024swg,Planas:2024vnq,Iglesias:2025tnt,Lousto:2022hoq} for related efforts in this direction).

While efficient waveform generation frameworks (such as the FastEMRIWaveforms (FEW) package~\cite{chapman_bird_2025_15630565,FEW}) have been developed to make data analysis computationally feasible~\cite{Chua:2020stf,Katz:2021yft,Speri:2023jte,Chapman-Bird:2025xtd,Strusberg:2025qfv}, generating purely numerical GSF waveforms across the high-dimensional EMRI parameter space remains prohibitively expensive even at the adiabatic order. This is particularly true in the early inspiral phase, where the orbital separation can be arbitrarily large, and for generic orbital configurations. Consequently, fast analytical post-Newtonian (PN) expansions are required to complement numerical GSF data by efficiently covering this expansive parameter space~\cite{Skoupy:2024jsi,Honet:2025gge,Trestini:2026tky}. These analytical inputs also provide crucial calibrations for various binary inspiral template models, such as the Effective-One-Body (EOB) framework~\cite{Albertini:2025wez,vandeMeent:2017bcc,Nishimura:2026nse} and the phenomenological (Phenom) family~\cite{Pratten:2020fqn} (see, e.g., Refs.~\cite{Isoyama:2020lls,LISAConsortiumWaveformWorkingGroup:2023arg} for reviews).

The development of analytical PN expansions has a rich history, with traditional approaches achieving foundational successes in modelling the dynamics of arbitrary-mass binaries~\cite{Blanchet:2013haa,Poisson_Will_book,Porto:2016pyg,Levi:2018nxp,Schafer:2018jfw}.
In the extreme mass-ratio limit, these efforts are powerfully complemented by deriving high-order PN expansions directly from BHPT~\cite{Mino:1997bx,Sasaki:2003xr,Munna:2020civ}. 
While analytical fluxes for specialized configurations, such as circular or equatorial orbits, have been calculated to very high PN orders, e.g.,  Refs.~\cite{Fujita:2011zk,Fujita:2012cm,Forseth:2015oua,Fujita:2014eta,Sago:2024mgh,Munna:2020iju,Munna:2020juq,Castillo:2024isq,Castillo:2025ljw,Castillo:2025ryt,UNC-GR}, extending these formulas to generic bound orbits in Kerr spacetime is non-trivial due to the algebraic complexity of orbital motions, notably parametrized by the Carter constant~\cite{Carter:1968ks}: a unique characteristic of eccentric and precessing Kerr geodesics. 

Previous efforts derived GW fluxes up to 4PN order with the orbital eccentricity, $e$-expansion up to $O(e^6)$ for generic orbits~\cite{Sago:2015rpa} (built on earlier 2PN efforts~\cite{Sago:2005fn,Ganz:2007rf}), later extending to 5PN and ${O}(e^{10})$~\cite{Fujita:2020zxe,Isoyama:2021jjd}. 
Although calculating to arbitrarily high orders is theoretically possible, it presents two major practical challenges. First, the exponential growth in the number of terms results in analytical expressions so lengthy that evaluating them negates the computational speed advantage of an analytical model. 
Second, because standard PN expansions are asymptotic, adding higher-order terms yields diminishing returns---or even non-monotonic behaviour---in strong-field, highly eccentric regimes. Identifying a pragmatic truncation order is therefore essential.

In this paper, we establish the 6PN order as this pragmatic frontier. We present analytical formulas for the secular evolution of orbital parameters for generic bound orbits in Kerr spacetime, extending the state of the art to the 6PN order and including terms up to ${O}(e^{16})$ at linear order in the mass ratio.  
Building on established methodologies in the GSF theory~\cite{Sasaki:2003xr}, our results are exact in the inclination parameter and valid for arbitrary BH spins. 
Rather than expecting naive high-order PN truncations to perfectly describe the strong-field regime, we use these formulas to rigorously quantify how finite eccentricity affects both achievable accuracy and asymptotic PN convergence. To address the computational cost of evaluating these lengthy expressions, we construct an efficient ``hybrid'' approximation that combines different PN and eccentricity truncations, and we assess the performance of exponential resummation at these higher orders. 

By making the full expressions publicly available, these results are designed to integrate with community-driven repositories like the Black Hole Perturbation Toolkit~\cite{BHPT,wardell_2025_15969633} and Black Hole Perturbation Club (BHPC)~\cite{BHPC}, providing practical building blocks for accurate and fast waveform models of asymmetric-mass inspirals~\cite{BHPT_WaSABI,FEW}.

This paper is organized as follows.
Sec.~\ref{sec:overview} outlines the theoretical framework: Kerr geodesic motion and orbital parametrizations, the Teukolsky formalism and the flux-balance formula used to obtain GW fluxes.  
In Sec.~\ref{sec:results}, we compare the 6PN $O(e^{16})$ formulas with numerical results and examine convergence with PN order and eccentricity order.
In Sec.~\ref{sec:attempts}, we propose and test a hybrid model and evaluate exponential resummation.
Sec.~\ref{sec:summary} summarizes our results.

Throughout this paper we work in geometrized units with $G=c=1$ and adopt the metric signature $(-,+,+,+)$. Greek indices $\mu,\,\nu,\,\alpha,\,\beta,\,\dots$ denote coordinate components with respect to the Boyer--Lindquist coordinates $(t,\,r,\,\theta,\,\varphi)$, 
in which the Kerr metric takes the form
\begin{equation}
g_{\mu\nu}dx^\mu dx^\nu 
=
-\frac{\Delta}{\Sigma} \left[ d t - a\sin^2\theta\,d\varphi \right]^2
+\frac{\Sigma}{\Delta}\,dr^2 + \Sigma\,d\theta^2
+\frac{\sin^2\theta}{\Sigma} \left[(r^2+a^2)\,d\varphi-a\,dt\right]^2 \,,
\label{eq:Kerr}
\end{equation}
where $\Sigma := r^2+a^2\cos^2\theta$ and $\Delta := r^2-2Mr+a^2$.
Our curvature conventions follow Wald~\cite{Wald:1984rg}.

\section{Brief review of formulation: adiabatic inspirals in the self-force theory}
\label{sec:overview}

In this section, we briefly review the basic concepts required to calculate the secular changes of the orbital parameters due to gravitational radiation: Kerr geodesic equations, the Teukolsky formalism, and flux-balance formulas. The formulation presented here draws heavily upon the comprehensive review by Sasaki and Tagoshi~\cite{Sasaki:2003xr}, as well as a series of foundational works developed by the BHPC~\cite{BHPC}, specifically Refs.~\cite{Sago:2005fn,Sago:2005gd,Ganz:2007rf,Fujita:2011zk,Sago:2015rpa,Isoyama:2018sib}. We refer the reader to these references for detailed derivations.

\subsection{Geodesic equations in Kerr spacetime}

Timelike geodesics of the Kerr geometry are completely integrable, and they are uniquely characterized by three constants of motion (up to initial conditions) associated with the Killing(-Yano) symmetry of Kerr spacetime, i.e., 
the Killing vectors $t^{\mu}$ and $\phi^{\mu}$,  
and the Killing-St\"ackel tensor $K^{\mu \nu}$~\cite{Walker:1970un}.
These constants are the specific energy $\hat E$, azimuthal angular momentum $\hat L$, 
and Carter constant ${\hat Q}$~\cite{Carter:1968rr} of a particle, 
\begin{equation}\label{def-P}
 \hat E := - t^{\mu} u_{\mu} \,, 
 \quad
 \hat L :=  \phi^{\mu} u_{\mu}\,, 
 \quad
 {\hat Q} := K^{\mu \nu} u_{\mu} u_{\nu}\,,
\end{equation}
by using the four velocity, $u^{\mu}=dx^{\mu}/d\tau$.
Explicitly, ${\hat E} = - u_{t}$, ${\hat L} = u_{\phi}$, and 
\begin{equation}\label{eq:def-Q}
{\hat Q} 
=
a^2 (1 - {\hat E}^2) \cos^2 \theta 
+ {\hat L}^2 \cot^2 \theta + u_\theta^2 
+ ( a {\hat E} - {\hat L})^2 \,.
\end{equation}
Another variable related to the Carter constant, 
\begin{equation}\label{eq:def-C}
{\hat C} := {\hat Q} - (a {\hat E} - {\hat L})^2 
=
a^2 (1 - {\hat E}^2) \cos^2 \theta 
+ {\hat L}^2 \cot^2 \theta + u_\theta^2 \,, 
\end{equation}
is used as one of the orbital parameters, and 
generic orbits in Kerr spacetime are parametrized by three constants of motion,
$\{\hat{E},\, \hat{L},\, \hat{C}\}$ in this paper 
(see Ref.~\cite{Fujita:2009bp} for the analytical solutions).
The energy, angular momentum, and Carter constant of a particle with mass $\mu$
denoted by $\{E,\, L,\, C\}$, are obtained by
\begin{equation}
E:= \mu\,\hat{E} \,, \quad
L:= \mu\,\hat{L} \,, \quad
C:= \mu^2\,\hat{C} \,.
\end{equation}

With the three constants of motion, $\{\hat{E},\, \hat{L},\, \hat{C}\}$, 
the equations of timelike geodesics in Kerr spacetime can be expressed as
\begin{align}
\left( \frac{dr}{d\lambda} \right)^2 =& R(r) \,, \quad
\left( \frac{d\theta}{d\lambda} \right)^2 = \Theta(\cos\theta) \,, 
\label{eq:geodesics_r_theta} \\
\frac{dt}{d\lambda} =& V_{tr}(r) + V_{t\theta}(\cos\theta) \,, \quad
\frac{d\varphi}{d\lambda} = V_{\varphi r}(r) + V_{\varphi\theta}(\cos\theta) \,,
\label{eq:geodesics_t_phi}
\end{align}
where $\lambda$ is the Carter-Mino time defined by $d\lambda = d\tau/\Sigma$, and using $z=\cos \theta$, 
\begin{align}
P(r) :=& \hat{E}(r^2+a^2) - a\hat{L} \,, 
\label{eq:P(r)} \\
R(r) :=&
\left[ P(r) \right]^2 - \Delta \left[ r^2 + (a\hat{E} - \hat{L})^2 + \hat{C} \right] \,, \\
\Theta(z) :=& \hat{C}
- \left[ \hat{C} + a^2(1-\hat{E}^2) + \hat{L}^2 \right] z^2
+ a^2 (1-\hat{E}^2) z^4 \,, \\
V_{tr}(r) :=& \frac{r^2+a^2}{\Delta} P(r) \,, \quad
V_{t\theta}(z) := -a ( a\hat{E} - \hat{L} ) + a^2 \hat{E} z^2 \,, \\
V_{\varphi r}(r) :=& \frac{a}{\Delta} P(r) \,, \quad
V_{\varphi\theta} (z) := - a\hat{E} + \frac{\hat{L}}{1-z^2} \,.
\end{align}

\subsection{Choice of orbital parameters}

There are various choices of the parametrization of an orbit in Kerr spacetime, instead of 
$\{\hat{E},\, \hat{L},\, \hat{C}\}$. 
One of choices is $\{r_\textrm{max},\, r_\textrm{min},\, \theta_\textrm{min}\}$, 
where $r_\textrm{max}$ and $r_\textrm{min}$ are the maximum and minimum values of the radial
coordinate on a bound orbit respectively, and $\theta_\textrm{min}$ is the minimum values
of the polar coordinate.
The set of $\{r_\textrm{max},\, r_\textrm{min}\}$ is often replaced by Darwin parameters $\{p,\, e\}$ defined by
\begin{equation}
p := \frac{2 r_\textrm{max} r_\textrm{min}}{M(r_\textrm{max} + r_\textrm{min})} \,,
\qquad
e := \frac{r_\textrm{max} - r_\textrm{min}}{r_\textrm{max} + r_\textrm{min}} \,.
\label{eq:def-pe}
\end{equation}
Following celestial mechanics, $p$ and $e$ are called the (dimensionless) semi-latus rectum
and eccentricity, respectively. 

For Kerr case, there is some ambiguity in the definition of inclination angle.
A useful definition of inclination angle, $\iota$, is
\begin{equation} \label{eq:def-iota}
 \cos\iota := \frac{\hat{L}}{\sqrt{\hat{L}^2 + \hat{C}}} \,.
\end{equation}
In our previous works for calculation of radiation reaction effect, we use $\cos\iota$
\footnote{We usually treat the notation, $Y := \cos\iota$, in our previous works, and use $Y$ to simplify the expression of the formulas in Sec.~\ref{sec:PN_formulas}.}
as an inclination parameter because it simplifies the analytic calculation.
On the other hand, in most numerical works, a different definition of
inclination angle, $\theta_\textrm{inc} := \pi/2-\textrm{sgn}(L)\,\theta_\textrm{min}$, is used.
Combining $\Theta(\cos\theta_\textrm{min})=0$ and Eq.~\eqref{eq:def-iota}, 
the relation between $\iota$ and $\theta_\textrm{inc}$ can be obtained as
\begin{equation} 
\cos\iota =
\cos\theta_\textrm{inc} \left[
1 + \frac{a^2(1 - \hat{E}^2)}{\hat{L}^2} \cos^2 \theta_\textrm{inc} \,
(1 - \cos^2 \theta_\textrm{inc}) 
\right]^{-1/2} \,.
\label{eq:rel-iota-thetainc}
\end{equation}
Obviously, $\theta_\mathrm{inc}$ coincides with $\iota$ for non-spinning (Schwarzschild) case.

For performing PN expansions, we introduce the velocity parameter 
$v := \sqrt{1/p}$, which corresponds to the typical magnitude of the orbital velocity.
Expanding a quantity with respect to $v$ and truncating after $j$-th order of $v$ gives its
$(j/2)$PN expression.
For example, up to the 3.5PN order, the relation in Eq.~\eqref{eq:rel-iota-thetainc} is given
in the PN form as
\begin{equation}
\cos\iota = \cos\theta_\textrm{inc} \left[ 1
- \frac{(1-e^2)(1-\cos^2\theta_\textrm{inc})q^2}{2} v^4 \left(
1 - 4v^2 + 8 q \cos\theta_\textrm{inc} v^3 + \cdots
\right) \right] \,,
\end{equation}
where $q := a/M$ is dimensionless, 
and our convention is that $q \hat{L} > 0$ ($q \hat{L} < 0$) represents prograde (retrograde) orbits.

\subsection{Fundamental frequencies}

For a bound orbit in Kerr geometry, there are two librational periods, in $\lambda$, 
related to the radial and polar motions obtained by
\begin{equation}
\Lambda_r := 2 \int_{r_\mathrm{min}}^{r_\mathrm{max}} \frac{dr}{\sqrt{R(r)}} \,,
\quad
\Lambda_\theta := 4 \int_{\theta_\mathrm{min}}^{\pi/2} \frac{d\theta}{\sqrt{\Theta(\cos\theta)}} \,,
\label{eq:libration_period}
\end{equation}
and the corresponding frequencies are defined as
\begin{equation}
\Upsilon_r := \frac{2\pi}{\Lambda_r} \,, \quad
\Upsilon_\theta := \frac{2\pi}{\Lambda_\theta} \,.
\label{eq:libration_frequency}
\end{equation}
With the librational frequencies, the radial and polar motions can be expressed in terms of the
Fourier series as
\begin{align}
r(\lambda) =& p M \sum_{n=0}^\infty \alpha_n \cos n\Upsilon_r \lambda \,,
\label{eq:r-motion} \\
\cos\theta(\lambda) =& \sin\iota \sum_{k=0}^\infty \beta_k \sin k\Upsilon_\theta \lambda \,,
\label{eq:theta-motion}
\end{align}
where we choose the initial values as $r(\lambda=0)=r_\mathrm{min}$ and $\theta(\lambda=0)=\pi/2$.
Since the Fourier coefficients of $r(\lambda)$ are given as $\alpha_n=O(e^n)$, the series can be
truncated at the finite number, $n_0$, if we consider the $e$-expansion up to $O(e^{n_0})$.
In a similar manner, the series of $\cos\theta(\lambda)$ can be truncated at the finite number, $k_0+1$,
to derive the $k_0$PN formula because $\beta_k=O(v^{2k-2})$ for odd $k$ and $\beta_k=0$ for even $k$.

Since the temporal and azimuthal components of the geodesic equations, 
Eq.~\eqref{eq:geodesics_t_phi}, are divided into two parts attributable to 
the radial and polar dependences, the temporal and azimuthal motions can be
expressed by
\begin{align}
t(\lambda) =& \Upsilon_t \lambda
+ \sum_{n=1}^\infty \tilde{t}_n^{(r)} \sin n\Upsilon_r \lambda
+ \sum_{k=1}^\infty \tilde{t}_k^{(\theta)} \sin k\Upsilon_\theta \lambda \,,
\label{eq:t-motion} \\
\varphi(\lambda) =& \Upsilon_\varphi \lambda
+ \sum_{n=1}^\infty \tilde{\varphi}_n^{(r)} \sin n\Upsilon_r \lambda
+ \sum_{k=1}^\infty \tilde{\varphi}_k^{(\theta)} \sin k\Upsilon_\theta \lambda \,,
\label{eq:phi-motion}
\end{align}
under the initial conditions of $t(\lambda=0)=0$ and $\varphi(\lambda=0)=0$.
Here, we introduce
\begin{equation}
\Upsilon_t := \left\langle \frac{dt}{d\lambda} \right\rangle_\lambda \,, \quad
\Upsilon_\varphi := \left\langle \frac{d\varphi}{d\lambda} \right\rangle_\lambda \,,
\end{equation}
with the time average of a function of $\lambda$, $X(\lambda)$, defined by
\begin{equation}
\left\langle X(\lambda) \right\rangle_\lambda :=
\lim_{T\to\infty} \frac{1}{2T} \int_{-T}^{T} X(\lambda) \,d\lambda \,.
\label{eq:ave_lambda}
\end{equation}
$\Upsilon_\varphi$ corresponds to the rotational frequency related to the azimuthal motion.
$\Upsilon_t$ is used to convert the average over $\lambda$ to the average over $t$ as~\cite{Drasco:2005is}
\begin{equation}
\left\langle \frac{dX}{dt} \right\rangle_t
= \Upsilon_t^{-1} \left\langle \frac{dX}{d\lambda} \right\rangle_\lambda \,.
\end{equation}
The frequencies with respect to $\lambda$ can be converted to those with respect 
to $t$, $\Omega_a$, as
\begin{equation}
\Omega_a = \frac{\Upsilon_a}{\Upsilon_t} \,, \quad (a=r,\, \theta,\, \varphi) \,.
\label{eq:Omega_a}
\end{equation}
$\Omega_a$ are called the fundamental frequencies, which characterize the frequency of 
GWs from EMRIs.

\subsection{Teukolsky formalism in Kerr spacetime: point particle source}

The Teukolsky formalism~\cite{Teukolsky:1973ha} is used to describe perturbations in Kerr spacetime, and the gravitational perturbations are described by the Weyl scalar, 
\begin{equation}
    \Psi_4 = \sum_{\ell m} \int d\omega\, R_{\Lambda}(r)\, S_{\Lambda}(\theta)\,{\rm e}^{im\varphi-i\omega t} \,,
\end{equation}
where $S_{\Lambda}(\theta)$ is the so-called spheroidal harmonics, satisfying the angular
Teukolsky equation and $\Lambda$ denotes a set of indices in the Fourier-harmonic
expansion, $\{\ell,\,m,\,\omega\}$.

The radial part of $\Psi_4$, $R_\Lambda(r)$, satisfies the radial Teukolsky equation,
\begin{equation}
\Delta^2 \frac{d}{dr} \left( \frac{1}{\Delta} \frac{dR_\Lambda(r)}{dr} \right)
+ \left[ \frac{K^2 + 4i(r-M)K}{\Delta} - 8i\omega r - \bar{\lambda} \right] R_\Lambda(r)
= T_\Lambda(r) \,,
\label{eq:radial_Teukolsky}
\end{equation}
where $K := (r^2+a^2)\omega-ma$ and $\bar{\lambda}$ is the eigenvalue of spheroidal harmonics. 
The source term, $T_\Lambda(r)$, is constructed with the energy-momentum
tensor of a point particle moving along an orbit described by the geodesic equations,
Eqs.~\eqref{eq:geodesics_r_theta} and \eqref{eq:geodesics_t_phi}
\footnote{The specific expression of $T_\Lambda(r)$ can be found in, e.g., the review in Sec.~2 of Ref.~\cite{Sasaki:2003xr}.}.

By using the Green's function method, the solution of Eq.~\eqref{eq:radial_Teukolsky} is obtained as
\begin{equation}
R_\Lambda(r) = \frac{1}{W(R_\Lambda^\mathrm{in},\, R_\Lambda^\mathrm{up})}
\left[ R_\Lambda^\mathrm{up}(r) \int_{r_+}^r \frac{R_\Lambda^\mathrm{in}(r')T_\Lambda(r')}{\Delta^2} dr'
+ R_\Lambda^\mathrm{in}(r) \int_{r}^\infty \frac{R_\Lambda^\mathrm{up}(r')T_\Lambda(r')}{\Delta^2} dr'
\right] \,,
\label{eq:solution_R}
\end{equation}
where $R_\Lambda^\mathrm{in}$ and $R_\Lambda^\mathrm{up}$ are two kinds of homogeneous solutions of
Eq.~\eqref{eq:radial_Teukolsky} which satisfy the asymptotic forms as
\begin{align}
  R_\Lambda^\mathrm{in} \to& \left\{
\begin{array}{ll}
B_\Lambda^{\mathrm{trans}} \Delta^2
\mathrm{e}^{-i \mathrm{k} r^*} \,,
& \, \text{for}\ r\to r_{+} \,, \\
r^3 B_\Lambda^{\mathrm{ref}} \mathrm{e}^{i\omega r^*}
+r^{-1}B_\Lambda^{\mathrm{inc}}
\mathrm{e}^{-i\omega r^*} \,, 
& \, \text{for}\ r\to+\infty \,, \\
\end{array}
\right.\cr
  R_\Lambda^\mathrm{up} \to& \left\{
\begin{array}{ll}
C_\Lambda^{\mathrm{up}} 
\mathrm{e}^{i \mathrm{k} r^*}
+ \Delta^2 C_\Lambda^{\mathrm{ref}} 
\mathrm{e}^{-i \mathrm{k} r^*} \,,
& \,\,\,\,\,\,\,\,\, \text{for}\ r\to r_{+} \,, \\
r^3 C_\Lambda^{\mathrm{trans}}
\mathrm{e}^{i\omega r^*} \,, 
& \,\,\,\,\,\,\,\,\, \text{for}\ r\to+\infty \,, \\
\end{array}
\right.
\label{eq:bc_Rinup}
\end{align}
with $\mathrm{k} := \omega-ma/(2Mr_+)$, $r_+:=M+\sqrt{M^2-a^2}$, and $r^*$ is the tortoise coordinate.
The Wronskian of the two homogeneous solutions is given by
\begin{equation}
W(R_\Lambda^\mathrm{in}, R_\Lambda^\mathrm{up}) = 
\frac{1}{\Delta} \left[ 
R_\Lambda^\mathrm{in}(r) \frac{d}{dr} R_\Lambda^\mathrm{up}(r)
- R_\Lambda^\mathrm{up}(r) \frac{d}{dr} R_\Lambda^\mathrm{in}(r)
\right] =
2 i \omega B_\Lambda^\mathrm{inc} C_\Lambda^\mathrm{trans} \,.
\label{eq:Wronskian}
\end{equation}
The analytic PN expressions for $R_\Lambda^\mathrm{in/up}$ and the coefficients of their 
asymptotic forms can be calculated by the systematic method developed by Mano, Suzuki, and 
Takasugi (MST) \cite{Mano:1996vt, Sasaki:2003xr}.

With Eqs.~\eqref{eq:bc_Rinup} and \eqref{eq:Wronskian}, the asymptotic forms of the radial
solution, Eq.~\eqref{eq:solution_R}, are found as
\begin{align}
R_\Lambda(r \to r_+) \to& \,\,
\frac{B_\Lambda^\mathrm{trans}\Delta^2
\mathrm{e}^{-i \mathrm{k} r^*}}
{2i\omega C_\Lambda^\mathrm{trans} B_\Lambda^\mathrm{inc}}
\int_{r_+}^\infty \frac{R_\Lambda^\mathrm{up}(r') T_\Lambda(r')}{\Delta^2} dr'
=:
\mu Z_\Lambda^\mathrm{H} \Delta 
\mathrm{e}^{-i \mathrm{k} r^*} \,, 
\label{eq:R_horizon} \\
R_\Lambda(r \to \infty) \to& \,\,
\frac{r^3 \mathrm{e}^{i\omega r^*}}{2i\omega B_\Lambda^\mathrm{inc}}
\int_{r_+}^\infty \frac{R_\Lambda^\mathrm{in}(r') T_\Lambda(r')}{\Delta^2} dr'
=:
\mu Z_\Lambda^\infty r^3 \mathrm{e}^{i\omega r^*} \,,
\label{eq:R_infinity}
\end{align}
where $Z_\Lambda^\mathrm{H}$ and $Z_\Lambda^\infty$ are the amplitudes of the partial
waves on the horizon and at infinity, respectively, which are necessary for calculating the secular
changes of the orbital parameters (see Eqs.~\eqref{eq:Edot}--\eqref{eq:Cdot} in the next subsection).
Performing the integrals in Eqs.~\eqref{eq:R_horizon} and \eqref{eq:R_infinity} analytically
is involved. It is usually simplified by using the expansions with respect to the eccentricity
and inclination in addition to the PN expansion (e.g., Refs.~\cite{Sago:2005fn,Sasaki:2003xr,Mano:1996vt}).
Since, in this work, we apply a method to calculate $Z_\Lambda^{\mathrm{H/}\infty}$ without the
aid of expansion of inclination developed in Ref.~\cite{Ganz:2007rf}, our analytic formulas can
be used for arbitrary inclination angle.

The frequency, $\omega$, becomes discrete in the case of a bound orbit, and the amplitudes are described by
\begin{equation}
    Z_\Lambda^{{\rm H}, \infty} = 2\,\pi\,\delta(\omega-\omega_{m k n})\, \tilde{Z}_{\tilde{\Lambda}}^{{\rm H}, \infty} \,.
    \label{eq:tildeZ}
\end{equation}
$\tilde{\Lambda}$ is the set of indices, $\{\ell,\,m,\,k,\,n\}$ 
and $\omega_{mkn}$ is written by using the fundamental frequencies given in Eq.~\eqref{eq:Omega_a} as
\begin{equation}
    \omega_{mkn} =  m\, \Omega_\varphi + k\, \Omega_\theta + n\, \Omega_r \,.
\end{equation}

\subsection{Flux balance formulas}

With the amplitudes of the partial waves, Eq.~\eqref{eq:tildeZ},
the secular changes of the orbital parameters,
$\{E,\, L,\, C\}$, are expressed as \cite{Sago:2005fn,Sago:2005gd} (see also Refs.~\cite{Isoyama:2018sib,Grant:2024ivt})
\begin{align}
\left\langle \frac{dE}{dt} \right\rangle_t
=&
- \mu^2 \sum_{\tilde\Lambda}
\frac{1}{4\pi\omega_{mkn}^2}
\left(
\left| \tilde{Z}_{\tilde\Lambda}^\infty \right|^2
+ \alpha_{\ell m}(\omega_{mkn})
\left| \tilde{Z}_{\tilde\Lambda}^{\rm H} \right|^2
\right) \,,
\label{eq:Edot} \\
\left\langle \frac{dL}{dt} \right\rangle_t
=&
- \mu^2 \sum_{\tilde\Lambda}
\frac{m}{4\pi\omega_{mkn}^3}
\left(
\left| \tilde{Z}_{\tilde\Lambda}^\infty \right|^2
+ \alpha_{\ell m}(\omega_{mkn})
\left| \tilde{Z}_{\tilde\Lambda}^{\rm H} \right|^2
\right) \,,
\label{eq:Ldot} \\
\left\langle \frac{dC}{dt} \right\rangle_t
=&
- 2\left\langle a^2 E \cos^2\theta \right\rangle_\lambda
\left\langle \frac{dE}{dt} \right\rangle_t
+ 2\left\langle L \cot^2\theta \right\rangle_\lambda
\left\langle \frac{dL}{dt} \right\rangle_t
\cr &
- \mu^3 \sum_{\tilde\Lambda}
\frac{k\, \Upsilon_\theta}
{2\pi\omega_{mkn}^3}
\left(
\left| \tilde{Z}_{\tilde\Lambda}^\infty \right|^2
+ \alpha_{\ell m}(\omega_{mkn})
\left| \tilde{Z}_{\tilde\Lambda}^{\rm H} \right|^2
\right) \,.
\label{eq:Cdot}
\end{align}
Here, 
\begin{equation}
\alpha_{\ell m}(\omega) =
\frac{256(2Mr_+)^5 \mathrm{k} (\mathrm{k}^2+4\tilde\epsilon^2)
(\mathrm{k}^2+16\tilde\epsilon^2) \omega^3}
{|{\cal C}_{S}(\omega)|^2} \,; \quad
\tilde\epsilon = \frac{\sqrt{M^2-a^2}}{4Mr_+} \,,
\end{equation}
and 
\begin{align}
|{\cal C}_{S}(\omega)|^2 =&
\left[ (\bar{\lambda}+2)^2 + 4a\omega m - 4a^2\omega^2 \right]
\left[ \bar{\lambda}^2 + 36a\omega m -36 a^2\omega^2 \right]
\cr &
+ (2\bar{\lambda}+3)(96a^2\omega^2 - 48a\omega m)
+ 144\omega^2(M^2-a^2) \,,
\end{align}
where ${\cal C}_{S}$ is the Starobinsky constant~\cite{Teukolsky:1974yv}.
As a result, the analytic formulas for a bound orbit can be obtained in the PN form of
\begin{equation}
\left\langle \frac{dI}{dt} \right\rangle_t =
\left(\frac{dI}{dt}\right)_\mathrm{N}
\sum_{j} \delta_I^{(j)} v^j \,,
\label{eq:delta_I_j}
\end{equation}
where $I=\{E,\, L,\, C\}$ and $( dI/dt )_\mathrm{N}$ are the leading contributions.
$\delta_I^{(j)}$ corresponds to the relative contribution of the $(j/2)$PN term, which depends on
$q$, $e$, $\iota$ (or $\theta_\mathrm{inc})$, and includes some terms with the power of $\ln v$
for $j \ge 6$. With the aid of the expansion with respect to $e$, $\delta_I^{(j)}$ is given
in the form of a power series in $e$, as an example shown in the next section.

\section{Results}
\label{sec:results}

\subsection{PN formulas with the expansion in eccentricity}
\label{sec:PN_formulas}

Based on the procedure presented in the previous section, we calculate 
Eqs.~\eqref{eq:Edot}, \eqref{eq:Ldot}, and \eqref{eq:Cdot} analytically in the PN expanded form.
The analytic formulas are given as truncated series in two variables, $v$ and $e$.
A formula truncated after $i$-th order of $v$ and $j$-th order of $e$ is called ``$(i/2)$PN $O(e^j)$ formula''.
For example, the analytic formula of $\langle dE/dt \rangle_t$
up to the 3PN $O(e^{16})$ are
\begin{equation}
\left\langle \frac{dE}{dt} \right\rangle_t =
\left(\frac{dE}{dt}\right)_\mathrm{N}
\sum_{j=0}^{6} \delta_E^{(j)} v^j \,,
\end{equation}
with
\begin{align*}
\delta_E^{(0)} =& 1 + \frac{73}{24} e^2 + \frac{37}{96} e^4, \\
\delta_E^{(1)} =& 0, \\
\delta_E^{(2)} =& -\frac{1247}{336} - \frac{9181}{672} e^2 + \frac{809}{128} e^4 + \frac{8609}{5376} e^6, \\
\delta_E^{(3)}=& 4 \pi-\frac{73}{12} Y q +\left(\frac{1375}{48} \pi-\frac{823}{24} Y q\right) e^{2}+\left(\frac{3935}{192} \pi-\frac{949}{32} Y q\right) e^{4}
\\&\hspace{1pt} 
+\left(\frac{10007}{9216} \pi-\frac{491}{192} Y q\right) e^{6}
+\frac{2321}{221184} \pi\,e^{8}-\frac{237857}{88473600} \pi  \,e^{10}+\frac{182863}{1061683200} \pi  \,e^{12}
\\&\hspace{1pt} 
+\frac{4987211}{1664719257600} \pi  \,e^{14}
-\frac{47839147}{8878502707200} \pi  \,e^{16},\\
\delta_E^{(4)}=& -\frac{44711}{9072}-\frac{329}{96} q^{2}+\frac{527}{96} Y^{2} q^{2}+\left(-\frac{172157}{2592}-\frac{4379}{192} q^{2}+\frac{6533}{192} Y^{2} q^{2}\right) e^{2}
\\&\hspace{1pt} 
+\left(-\frac{2764345}{24192}-\frac{3823}{256} q^{2}+\frac{6753}{256} Y^{2} q^{2}\right) e^{4}+\left(\frac{3743}{2304}-\frac{363}{512} q^{2}+\frac{2855}{1536} Y^{2} q^{2}\right) e^{6}
\\&\hspace{1pt} 
+\frac{198377}{64512} e^{8}-\frac{1645}{2048} e^{10}-\frac{3165}{8192} e^{12}-\frac{1825}{8192} e^{14}-\frac{4675}{32768} e^{16},\\
\delta_E^{(5)}=& -\frac{8191}{672} \pi+\frac{3665}{336} Y q-\frac{9}{32} Y \,q^{3}-\frac{15}{32} Y^{3} q^{3}
\\&\hspace{1pt} 
+\left(-\frac{44531}{336} \pi +\frac{827}{28} Y q -\frac{135}{64} Y \,q^{3}-\frac{225}{64} Y^{3} q^{3}\right) e^{2}
\\&\hspace{1pt} 
+\left(-\frac{4311389}{43008} \pi -\frac{113093}{1344} Y q -\frac{405}{256} Y \,q^{3}-\frac{675}{256} Y^{3} q^{3}\right) e^{4}
\\&\hspace{1pt} 
+\left(\frac{15670391}{387072} \pi -\frac{6215}{56} Y q -\frac{45}{512} Y \,q^{3}-\frac{75}{512} Y^{3} q^{3}\right) e^{6}
\\&\hspace{1pt} 
+\left(\frac{32271047}{7077888} \pi-\frac{144139}{14336} Y q\right) e^{8}+\left(\frac{9561889}{77414400} \pi+\frac{329}{256} Y q\right) e^{10}
\\&\hspace{1pt} 
+\left(-\frac{3228161479}{95126814720} \pi+\frac{633}{1024} Y q\right) e^{12}+\left(\frac{113251706317}{69918208819200} \pi+\frac{365}{1024} Y q\right) e^{14}
\\&\hspace{1pt} 
+\left(\frac{4364728552231}{35798122915430400} \pi+\frac{935}{4096} Y q\right) e^{16},\\
\delta_E^{(6)}=& \frac{16}{3} \pi^{2}-\frac{1712}{105} \gamma-\frac{3424}{105} \ln \! \left(2\right)+\frac{6643739519}{69854400}+\frac{135}{8} q^{2}-\frac{169}{6} \pi  Y q+\frac{73}{21} Y^{2} q^{2}
\\&\hspace{1pt} 
+\left(\frac{680}{9} \pi^{2}-\frac{13696}{315} \ln \! \left(2\right)-\frac{234009}{560} \ln \! \left(3\right)-\frac{14552}{63} \gamma+\frac{43072561991}{27941760}+\frac{205747}{1344} q^{2}
\right.\\&\hspace{24pt} \left.
-\frac{4339}{16} \pi  Y q+\frac{13697}{192} Y^{2} q^{2}\right) e^{2}
\\&\hspace{1pt} 
+\left(\frac{5171}{36} \pi^{2}-\frac{12295049}{1260} \ln \! \left(2\right)+\frac{2106081}{448} \ln \! \left(3\right)-\frac{553297}{1260} \gamma+\frac{919773569303}{279417600}
\right.\\&\hspace{24pt} \left.
+\frac{208571}{1792} q^{2}-\frac{42271}{96} \pi  Y q+\frac{471711}{1792} Y^{2} q^{2}\right) e^{4}
\\&\hspace{1pt} 
+\left(\frac{1751}{36} \pi^{2}-\frac{5224609375}{193536} \ln \! \left(5\right)+\frac{24908851}{252} \ln \! \left(2\right)-\frac{864819261}{35840} \ln \! \left(3\right)-\frac{187357}{1260} \gamma
\right.\\&\hspace{24pt} \left.
+\frac{308822406727}{186278400}+\frac{3253}{10752} q^{2}-\frac{4867907}{27648} \pi  Y q+\frac{289063}{1536} Y^{2} q^{2}\right) e^{6}
\\&\hspace{1pt} 
+\left(\frac{99}{64} \pi^{2}+\frac{496337890625}{1548288} \ln \! \left(5\right)-\frac{13091486273}{20160} \ln \! \left(2\right)-\frac{2480573403}{40960} \ln \! \left(3\right)
\right.\\&\hspace{24pt} \left.
-\frac{10593}{2240} \gamma+\frac{2087886792107}{2980454400}-\frac{12577}{86016} q^{2}-\frac{134279}{16384} \pi  Y q+\frac{1110551}{86016} Y^{2} q^{2}\right) e^{8}
\\&\hspace{1pt} 
+\left(-\frac{262029833984375}{148635648} \ln \! \left(5\right)+\frac{536743821479}{162000} \ln \! \left(2\right)+\frac{346080010951719}{229376000} \ln \! \left(3\right)
\right.\\&\hspace{24pt} \left.
-\frac{507989081563901}{884736000} \ln \! \left(7\right)+\frac{369490499473}{851558400}+\frac{329}{768} q^{2}-\frac{2289029}{17694720} \pi  Y q
\right.\\&\hspace{24pt} \left.
-\frac{1645}{1024} Y^{2} q^{2}\right) e^{10}
\\&\hspace{1pt} 
+\left(\frac{168770556640625}{28311552} \ln \! \left(5\right)-\frac{4242983254663}{216000} \ln \! \left(2\right)-\frac{1219799872597347}{131072000} \ln \! \left(3\right)
\right.\\&\hspace{24pt} \left.
+\frac{8635814386586317}{1179648000} \ln \! \left(7\right)+\frac{1726799683}{5529600}+\frac{211}{1024} q^{2}+\frac{4322471}{132710400} \pi  Y q
\right.\\&\hspace{24pt} \left.
-\frac{3165}{4096} Y^{2} q^{2}\right) e^{12}
\\&\hspace{1pt} 
+\left(-\frac{14561181430712890625}{1048773132288} \ln \! \left(5\right)+\frac{55686033875764823}{444528000} \ln \! \left(2\right)
\right.\\&\hspace{24pt} \left.
+\frac{6720555668600328741}{359661568000} \ln \! \left(3\right)-\frac{66791928411266395183}{1528823808000} \ln \! \left(7\right)+\frac{105621449}{423360}
\right.\\&\hspace{24pt} \left.
+\frac{365}{3072} q^{2}-\frac{68755189}{66588770304} \pi  Y q-\frac{1825}{4096} Y^{2} q^{2}\right) e^{14}
\\&\hspace{1pt} 
+\left(-\frac{4345447067236328125}{16780370116608} \ln \! \left(5\right)-\frac{8274561002983367327}{12802406400} \ln \! \left(2\right)
\right.\\&\hspace{24pt} \left.
+\frac{19844159570399695203}{164416716800} \ln \! \left(3\right)+\frac{52933986266203175903}{326149079040} \ln \! \left(7\right)
\right.\\&\hspace{24pt} \left.
+\frac{288319630969}{1387266048}+\frac{935}{12288} q^{2}+\frac{13941221843}{106542032486400} \pi  Y q-\frac{4675}{16384} Y^{2} q^{2}\right) e^{16}
\\&\hspace{1pt} 
+
\left(-\frac{1712}{105}-\frac{14552}{63} e^{2}-\frac{553297}{1260} e^{4}-\frac{187357}{1260} e^{6}-\frac{10593}{2240} e^{8}\right) \ln v,
\end{align*}
where $\gamma$ is the Euler constant, $Y=\cos\iota$, and  
the leading contribution is given by
\[
\left( \frac{dE}{dt} \right)_\mathrm{N} =
-\frac{32}{5} \left( \frac{\mu}{M} \right)^2 v^{10} \left( 1-e^2 \right)^{3/2} \,.
\]
In this work, we derived the 6PN $O(e^{16})$ formulas of the related variables 
(constants of motion, fundamental frequencies, and so on), in addition to $\langle dI/dt \rangle_t$.
Because their expressions are too lengthy and redundant to present in the paper, we publish
the raw data online~\cite{BHPC}.

\subsection{Comparison with numerical fluxes}

\begin{figure}[!t]
\includegraphics[bb=0 0 841 576, width=0.95\linewidth]{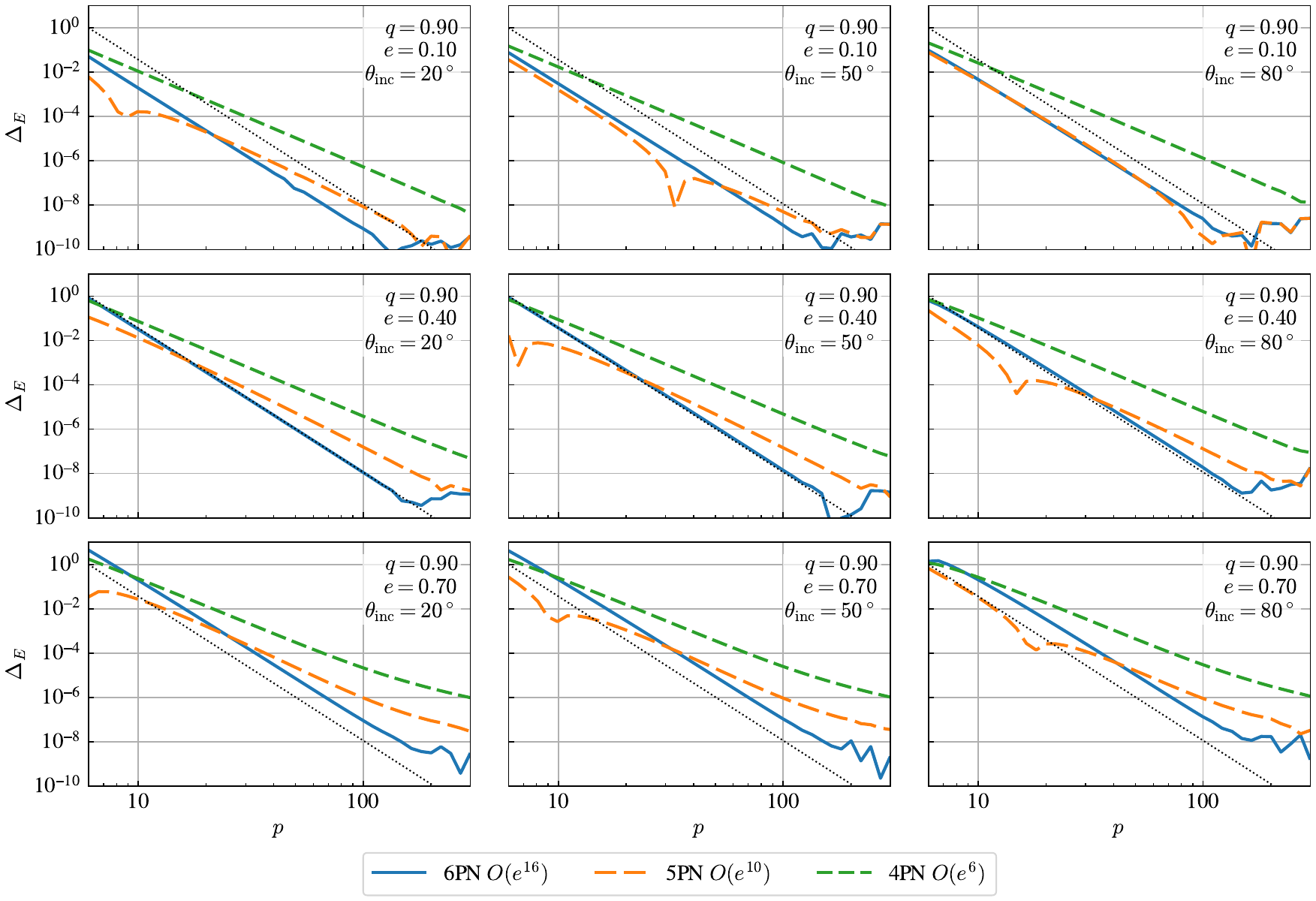}
\caption{
Relative errors in the analytic PN formulas for the secular change of the particle's
energy $E$ as a function of the (dimensionless) semi-latus rectum $p$ for the dimensionless 
spin parameter $q=0.90$ (see Eq.~\eqref{eq:relative_error}). 
We truncate the plots at $p=6$.
These plots correspond to Fig.~1 of Ref.~\cite{Sago:2015rpa}.
The dotted line in each plot is reference proportional to $1/p^{13/2}=v^{13}$.
The cases with the eccentricity $e=0.10$, $0.40$, and $0.70$ are shown from the top to the bottom panels,
and those with the inclination $\theta_\textrm{inc}=20^\circ$, $50^\circ$, and $80^\circ$ are shown from the left to the right panels.}
\label{fig:DelE_q09}
\end{figure}

\begin{figure}[!ht]
\centering
\includegraphics[bb=0 0 396 304, width=0.6\linewidth]{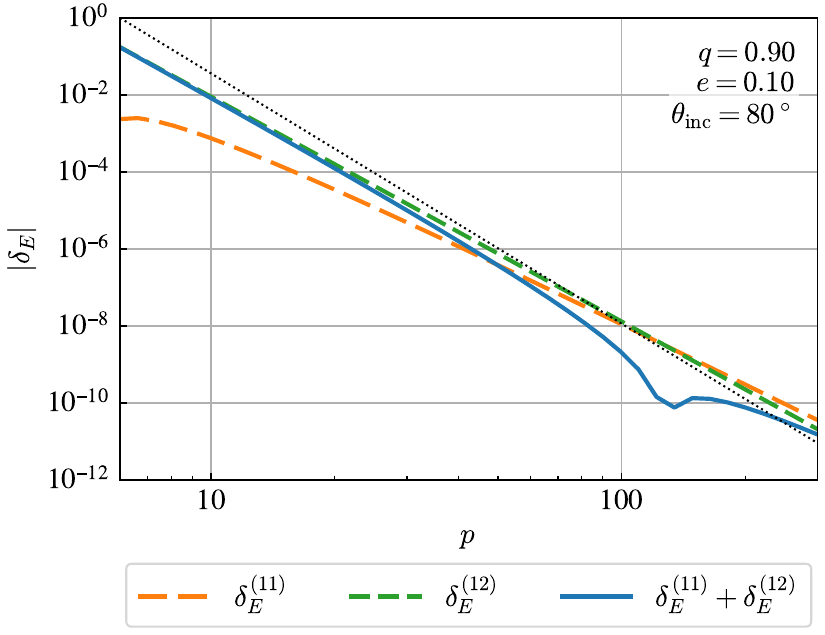}
\caption{Relative contributions of the 5.5PN and 6PN terms in the analytic formula for
$\langle dE/dt \rangle_t$ in the case of $q=0.90$, $e=0.10$, and $\theta_\mathrm{inc}=80^\circ$, corresponding the top right panel of Fig.~\ref{fig:DelE_q09}. $\delta_E^{(j)}$ is defined by Eq.~\eqref{eq:delta_I_j}.}
\label{fig:DelE_5PN6PN_q09e01inc80}
\end{figure}

To assess the accuracy of our analytic formulas of $\langle dI/dt \rangle_t$, we introduce
the relative error to the numerical results as
\begin{equation}
\Delta_{I} :=
\left|
1 - \left\langle\frac{dI}{dt}\right\rangle_t^{\rm Ana}
\bigg/ \left\langle\frac{dI}{dt}\right\rangle_t^{\rm Num} \right| \,.
\label{eq:relative_error}
\end{equation}
Here, $\left\langle{dI}/{dt}\right\rangle_t^{\rm Ana}$ denotes the analytic formula derived in this paper, and  $\left\langle{dI}/{dt}\right\rangle_t^{\rm Num}$ is the numerical result given by the method in Refs.~\cite{Fujita:2004rb,Fujita:2005kng,Fujita:2009us,Fujita:2020zxe}.
For computing $\left\langle{dI}/{dt}\right\rangle_t^{\rm Num}$, we must truncate the summation over $\tilde\Lambda=\{\ell,\,m,\,k,\,n\}$ in Eqs.~\eqref{eq:Edot}, \eqref{eq:Ldot}, and \eqref{eq:Cdot}. For the summation over $k$ and $n$, we set the relative error in the summation to be $10^{-8}$. For the summation over $m$, we compute all the $m$-mode from $-\ell$ to $\ell$. For the summation over $\ell$, to save computation time we set the maximum value of $\ell$, $\ell_{\rm max}$, as $\ell_{\rm max}=12$, while we stop the summation if the relative error in the summation becomes smaller than $10^{-8}$ for $\ell<\ell_{\rm max}$. 

Fig.~\ref{fig:DelE_q09} shows the plots of $\Delta_E$ as functions of $p$ for the cases with
$e=\{0.10,\, 0.40,\, 0.70\}$ and $\theta_\mathrm{inc}=\{20^\circ,\, 50^\circ,\, 80^\circ\}$. The dimensionless
spin is fixed as $q=0.90$. In the plots, one can see that the relative errors of the 6PN $O(e^{16})$
formula (blue solid lines) show a trend proportional to $p^{-13/2}$ as expected.
For the comparison, we also show $\Delta_E$ for the 4PN $O(e^6)$ formula
derived in Ref.~\cite{Sago:2015rpa} (green short-dashed lines), the 5PN $O(e^{10})$ formula used in
Ref.~\cite{Isoyama:2021jjd, Fujita:2020zxe} (orange, long-dashed lines) on the same panels.
For large $p$, the 6PN $O(e^{16})$ formula improves the accuracy compared to the lower PN formulas,
while it does not for small $p$. The 6PN $O(e^{16})$ formula seems to be even worse than the 5PN $O(e^{10})$.
It is known that the PN expansion does not converge monotonously as the expansion order increases,
although it tends to converge on average
(e.g., Refs.~\cite{Sago:2016xsp, Fujita:2017wjq, Munna:2020iju}). 
This behavior reflects the asymptotic nature of the PN expansion, where the inclusion of 
higher-order terms does not guarantee monotonic convergence at small orbital separations.

It should be noted that the relative error of the 5PN $O(e^{10})$ shows a similar
behaviour to that of the 6PN $O(e^{16})$ for $(e,\,\theta_\mathrm{inc})=(0.10,\, 80^\circ)$
case (top right panel of Fig.~\ref{fig:DelE_q09}). This behaviour is explained as follows:
The difference between the relative errors of the 5PN $O(e^{10})$ and 6PN $O(e^{16})$
is almost equal to the sum of the relative contributions from the 5.5PN and 6PN terms,
$\delta_E^{(11)}+\delta_E^{(12)}$
(now we can neglect corrections in the eccentricity expansion higher than $O(e^{10})$ because of the small
eccentricity, $e=0.10$).
In the left panel of Fig.~\ref{fig:DelE_5PN6PN_q09e01inc80}, 
we show $\delta_E^{(11)}$, $\delta_E^{(12)}$ and the
sum of them. One can find that $\delta_E^{(11)}$ and $\delta_E^{(12)}$ are proportional
to $p^{-11/2}$ and $p^{-6}$ respectively, while the sum shows a similar behaviour to
$p^{-13/2}$ unexpectedly (here we should note that $\delta_E^{(11)}$ is negative,
while $\delta_E^{(12)}$ is positive). As a result, the accuracy of the 5PN $O(e^{10})$
shows a similar behaviour to the 6PN $O(e^{16})$ for $(e,\,\theta_\mathrm{inc})=(0.10,\, 80^\circ$).
This coincidental cancellation of the 5.5PN and 6PN coefficients for this specific orbital 
geometry artificially suppresses the error of the 5PN truncation.

\begin{figure}[!t]
\includegraphics[bb=0 0 841 738, width=0.95\linewidth]{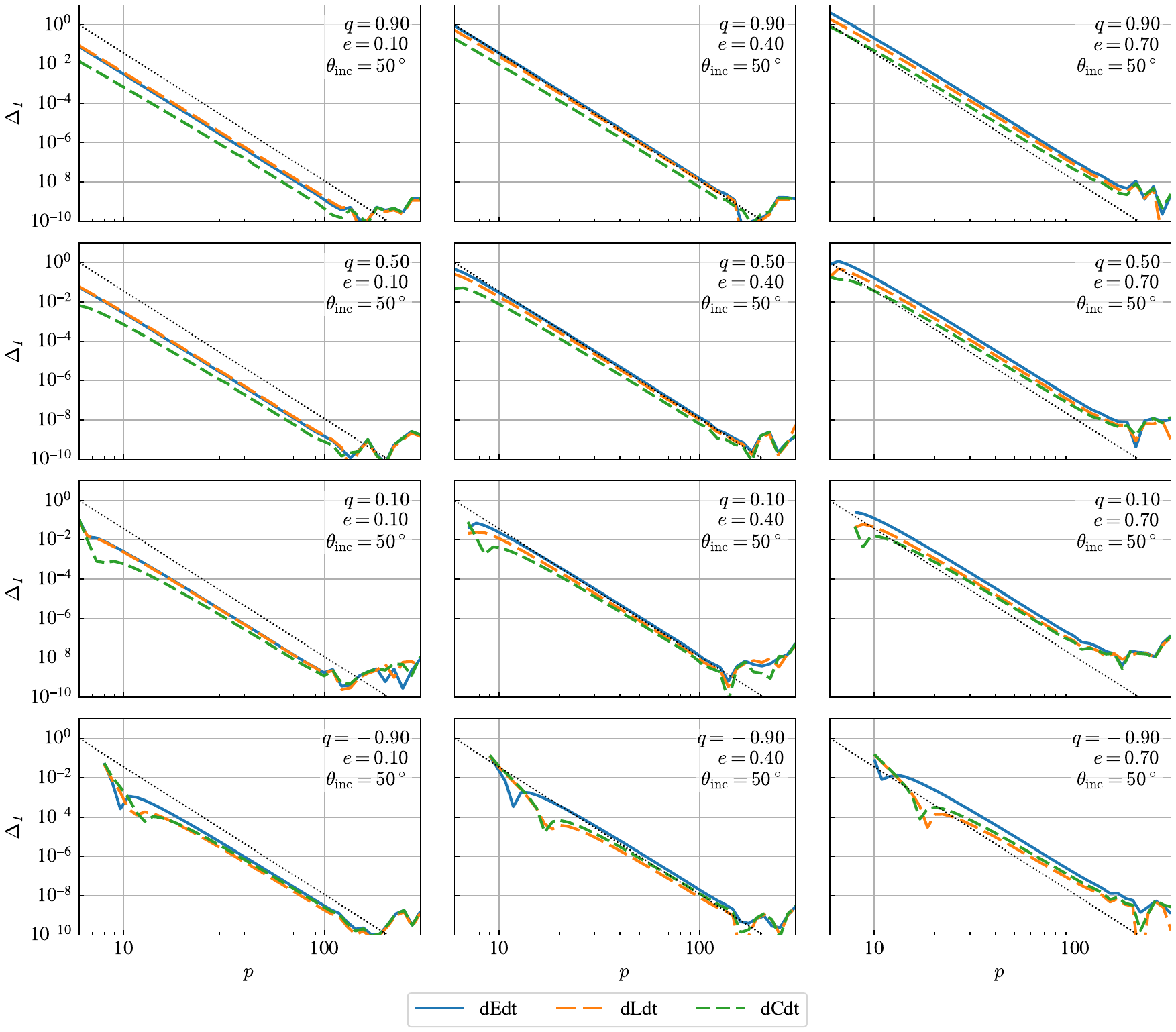}
\caption{Relative errors in the 6PN $O(e^{16})$ formulas for
$\langle dI/dt \rangle_t$ ($I=\{E,\, L,\, C\}$)
as a function of $p$ for $\theta_\textrm{inc}=50^\circ$  (see Eq.~\eqref{eq:relative_error}).
The cases with the dimensionless spin parameter $q=0.90$, $0.50$, $0.10$ and $-0.90$ are shown from the top to the bottom panels,
and those with the eccentricity $e=0.10$, $0.40$, and $0.70$ are shown from the left to the right panels.
This corresponds to Fig.~2 in Ref.~\cite{Sago:2015rpa}.}
\label{fig:DelI_inc50}
\end{figure}

Fig.~\ref{fig:DelI_inc50} shows the relative errors in the 6PN $O(e^{16})$ formulas for the
secular changes of $\{E,\, L,\, C\}$ as functions of $p$ for the cases with
$q=\{0.90,\, 0.50,\, 0.10,\, -0.90\}$ and $e=\{0.10,\, 0.40,\, 0.70\}$. In all cases, the inclination
angle is fixed as $\theta_\mathrm{inc}=50^\circ$. From the plots, one can find that the relative
errors, $\Delta_I$, for $I=\{E,\, L,\, C\}$ show a similar behaviour proportional
to $p^{-13/2}$. As mentioned in Ref.~\cite{Sago:2015rpa}, this fact suggests that investigating
the convergence of $\langle dE/dt \rangle_t$ is sufficient to see the performance of the PN
formulas for the secular changes of the orbital parameters.

\subsection{Impact of the higher order correction of the expansion with $e$}

\begin{figure}[!ht]
\includegraphics[bb=0 0 841 576, width=0.95\linewidth]{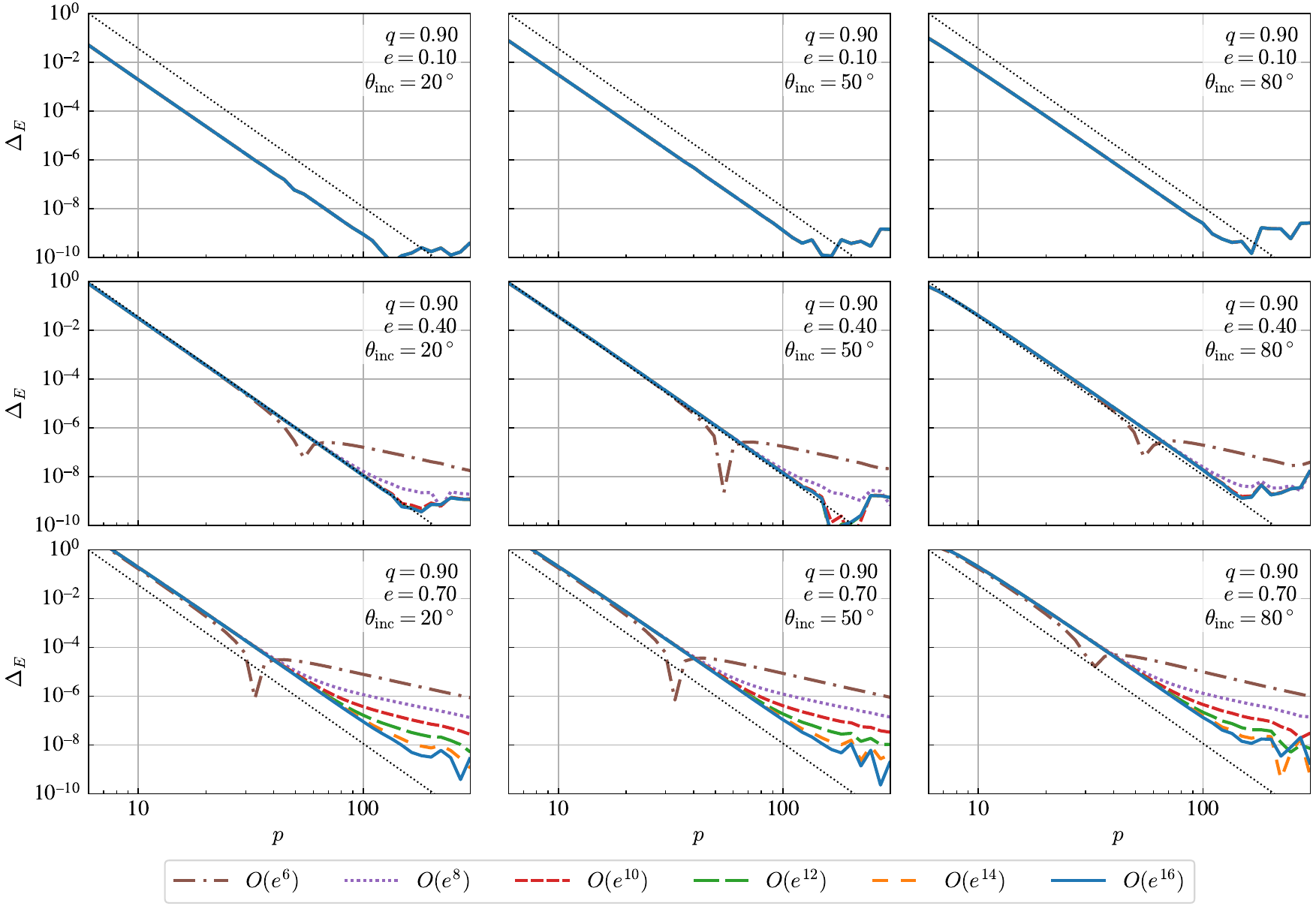}
\caption{Relative errors in the analytic 6PN formula for $\langle dE/dt \rangle_t$
with different order of expansion with respect to eccentricity.
The cases with the eccentricity $e=0.10$, $0.40$, and $0.70$ are shown from the top to the bottom panels,
and those with the inclination angle $\theta_\mathrm{inc}=20^\circ$, $50^\circ$, and $80^\circ$
are shown from the left to the right panels.
We fix $q=0.90$ for all plots.}
\label{fig:DelE_ne_q09}
\end{figure}

To see the impact of the higher order correction of the $e$-expansion
we compare the relative errors of the analytic formulas for $\langle dE/dt \rangle_t$
with different order corrections of $e$-expansion at the fixed 6PN order.
Fig.~\ref{fig:DelE_ne_q09} shows $\Delta_E$ for
the 6PN formula with $O(e^6)$, $O(e^8)$, $O(e^{10})$, $O(e^{12})$, $O(e^{14})$, and $O(e^{16})$
corrections. One can find that the $O(e^6)$ formula is sufficient to obtain the 6PN
accuracy in all range of $p$ shown in the plots in $e=0.10$ cases.
As the eccentricity is larger (for $e=0.40$, $0.70$ in the figure), the effect of the
higher order corrections becomes more apparent for large $p$ region.
For $e=0.70$, including at least up to the $O(e^{14})$ corrections is required to achieve the expected 6PN accuracy for $p \gtrsim 40$.
On the other hand, for $p \lesssim 40$, even the $O(e^8)$ formula achieves the 6PN
accuracy in $e=0.70$ cases. This is because the lower $e$-corrections in the higher
order terms than 6PN are more dominant for small $p$ region.
This fact leads us to produce a hybrid model combining the lower PN with higher
order of $e$ and the higher PN with lower order of $e$. 
We will discuss the potentiality of a hybrid model in Sec.~\ref{sec:hybrid}.

\begin{figure}[!ht]
\includegraphics[bb=0 0 856 597, width=0.95\linewidth]{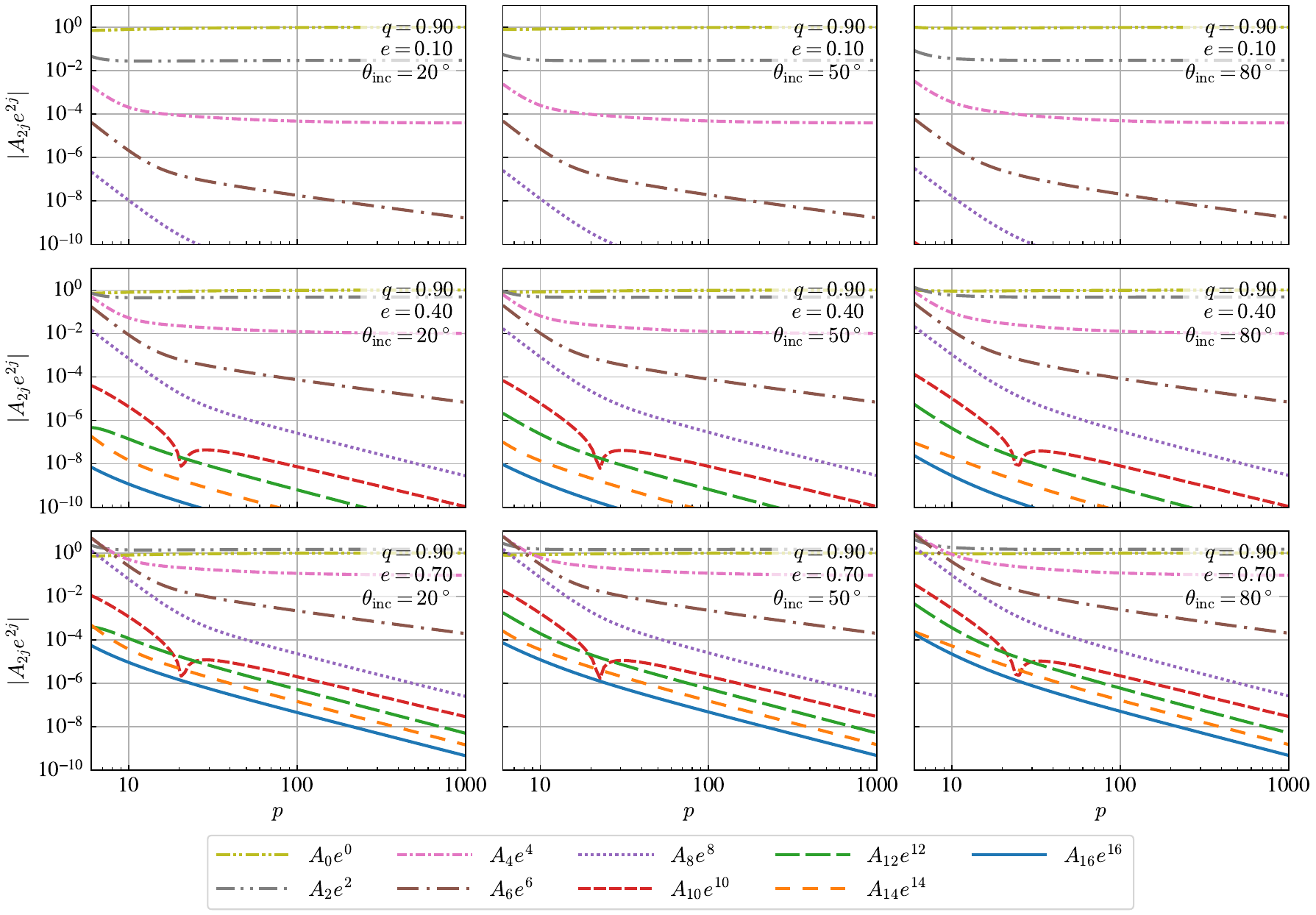}
\caption{Relative contribution of the $O(e^j)$ term in the PN formula for
$\langle dE/dt \rangle_t$ as a function of $p$ (see Eq.~\eqref{eq:Edot_e_exp}). 
The cases with the eccentricity $e=0.10$, $0.40$, and $0.70$ are shown from the top to the bottom panels,
and those with the inclination angle $\theta_\mathrm{inc}=20^\circ$, $50^\circ$, and $80^\circ$
are shown from the left to the right panels.
We fix $q=0.90$ in all plots.}
\label{fig:DelE_Anen_q09}
\end{figure}

The PN formula of $\langle dE/dt \rangle_t$ is expressed in the form of a Taylor series
with respect to $e$ as
\begin{equation}
\left\langle \frac{dE}{dt} \right\rangle_t^\mathrm{Ana} =
\left( \frac{dE}{dt} \right)_\mathrm{N}
\sum_{j=0}^{\infty} A_{2j}\, e^{2j} \,,
\label{eq:Edot_e_exp}
\end{equation}
where $A_{2j}$ are given in the form of the PN expansion and depend on $q$ and
$\theta_\mathrm{inc}$. The term $A_{2j}\, e^{2j}$ corresponds to the relative contribution
of the $O(e^{2j})$ term to the PN formula.
The leading term $A_0$ coincides with the value for the spherical orbit case.
Fig.~\ref{fig:DelE_Anen_q09} shows $A_{2j}\, e^{2j}$ as functions of $p$ for the cases with
$e=\{0.10,\, 0.40,\, 0.70\}$, $\theta_\mathrm{inc}=\{20^\circ,\, 50^\circ,\, 80^\circ\}$, and the
fixed $q=0.90$. For small eccentricity, the higher order terms have little contribution
to the total amount of $\langle dE/dt \rangle_t$. In the case of $e=0.10$ (shown in the
top panels of Fig.~\ref{fig:DelE_Anen_q09}), the contribution of the higher order terms
than $O(e^8)$ is less than $O(10^{-10})$. For large eccentricity, as a natural consequence,
the higher order contribution begins to affect the accuracy of the analytic formula.
For example, the contribution of $O(e^8)$--$O(e^{16})$ in large $p$ region for $e=0.70$
(the bottom panels of Fig.~\ref{fig:DelE_Anen_q09}) appears as the deviation from the
6.5PN slope shown in the bottom panels of Fig.~\ref{fig:DelE_ne_q09}.

\begin{figure}[!ht]
\includegraphics[bb=0 0 856 595, width=0.95\linewidth]{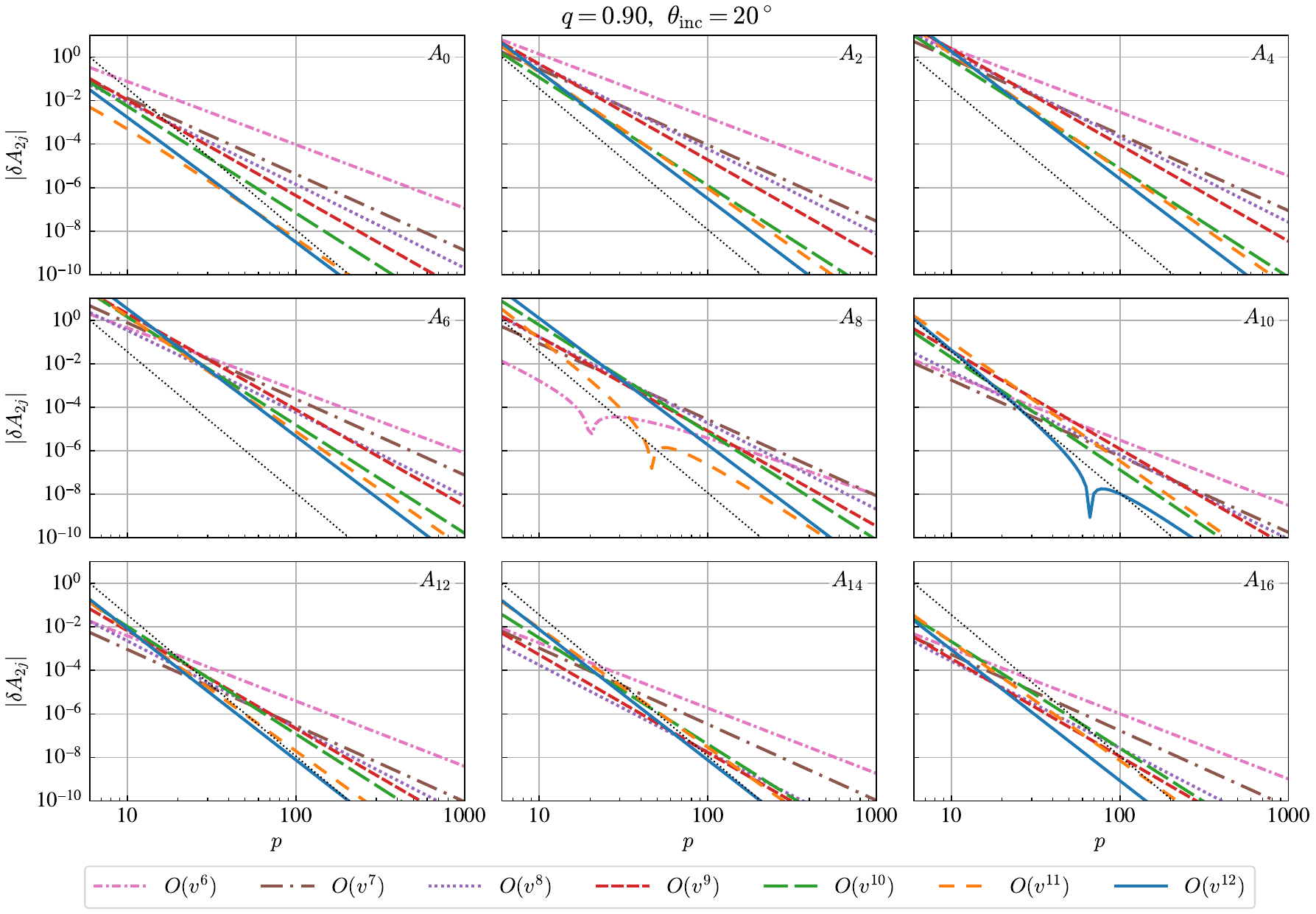}
\caption{Contribution of each PN term in $A_{2j}$ (see Eq.~\eqref{eq:delta_i_a_2j}) for the case with
$(q,\, \theta_\mathrm{inc})=(0.90,\, 20^\circ)$.}
\label{fig:delnA2j_q09_inc20}
\end{figure}

\begin{figure}[!ht]
\includegraphics[bb=0 0 856 595, width=0.95\linewidth]{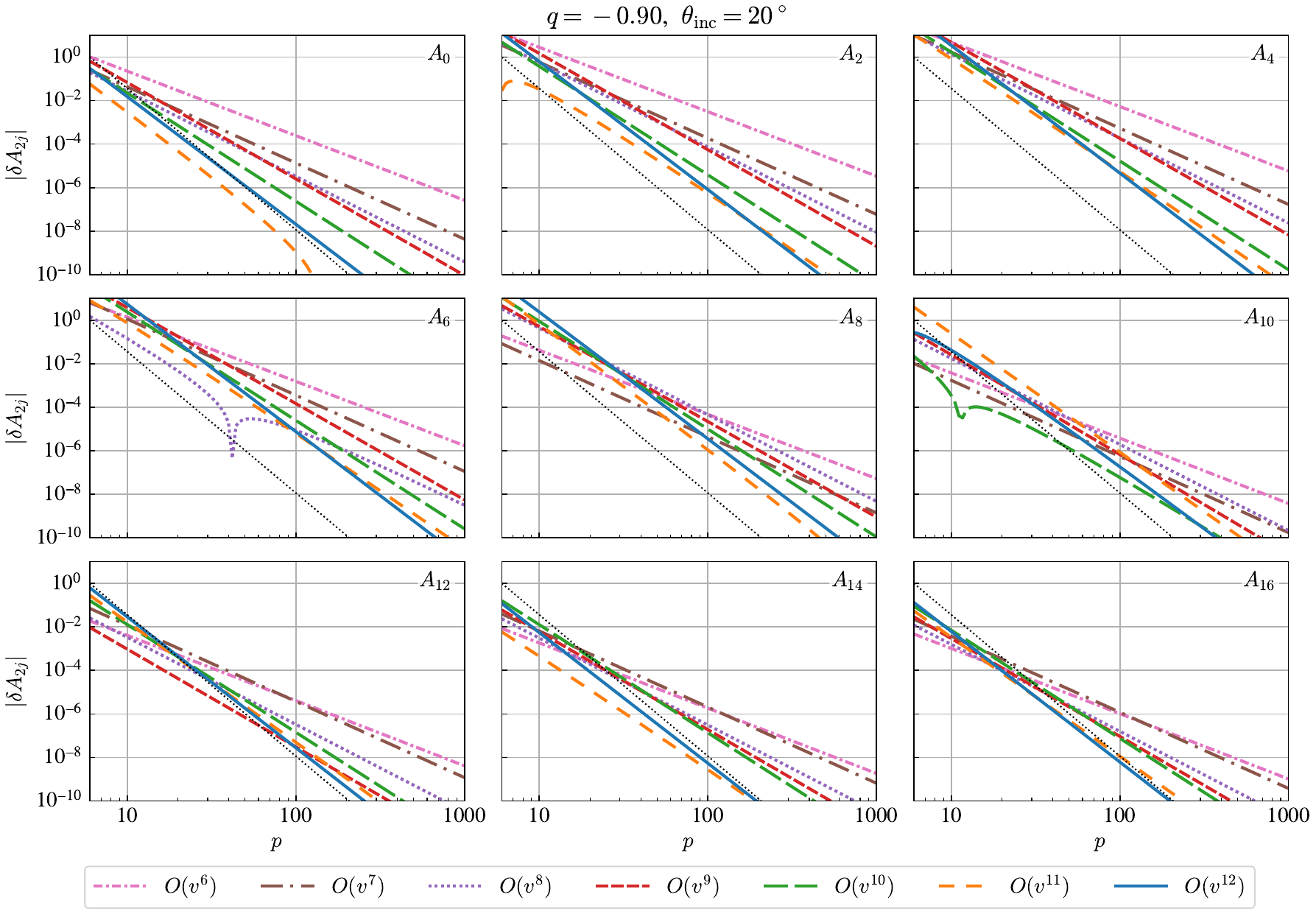}
\caption{
Same as Fig.~\ref{fig:delnA2j_q09_inc20}, but for 
$(q,\, \theta_\mathrm{inc})=(-0.90,\, 20^\circ)$.}
\label{fig:delnA2j_q-09_inc20}
\end{figure}

To investigate the impact of eccentricity on the PN convergence of the analytic formula
in more detail, we evaluate the contribution of each PN term in the coefficients
of the $e$-expansion, $A_{2j}$, defined as
\begin{equation}
\delta^{(i)} A_{2j} := A_{2j}^{(i)} - A_{2j}^{(i-1)} \,,
\label{eq:delta_i_a_2j}
\end{equation}
where $A_{2j}^{(i)}$ means the $(i/2)$PN formula of $A_{2j}$.
Fig.~\ref{fig:delnA2j_q09_inc20} shows $\delta^{(i)} A_{2j}$ for the case with
$(q,\, \theta_\mathrm{inc})=(0.90,\, 20^\circ)$.
For $p\gtrsim 100$, one can find $\delta^{(i)} A_{2j}$ decreases as $i$ increases for 
any $2j$ shown in the figure. This means that the PN formulas converge well for large
$p$ as expected.
On the other hand, the order of magnitude of the PN terms get messed up,
that is, the convergence of the PN formulas gets worse for smaller $p$.
Compared to the coefficients of the lower order with respect to $e$ (top panels in
Fig.~\ref{fig:delnA2j_q09_inc20}), the PN convergence of the higher order with respect to $e$ is worse
(middle and bottom panels). 
Similar behaviour is observed in different values of $(q,\, \theta_\mathrm{inc})$.
Fig.~\ref{fig:delnA2j_q-09_inc20} shows the case of
$(q,\, \theta_\mathrm{inc})=(-0.90,\, 20^\circ)$ for example.
This fact suggests that the accuracy of the analytic PN
formulas decreases when the orbital eccentricity becomes larger. This tendency is
consistent with the dependence of the relative errors in the PN formulas on $e$
(e.g., Figs~\ref{fig:DelE_q09} and \ref{fig:DelI_inc50}).

\begin{figure}[!ht]
\centering
\includegraphics[bb=0 0 395 323, width=0.6\linewidth]{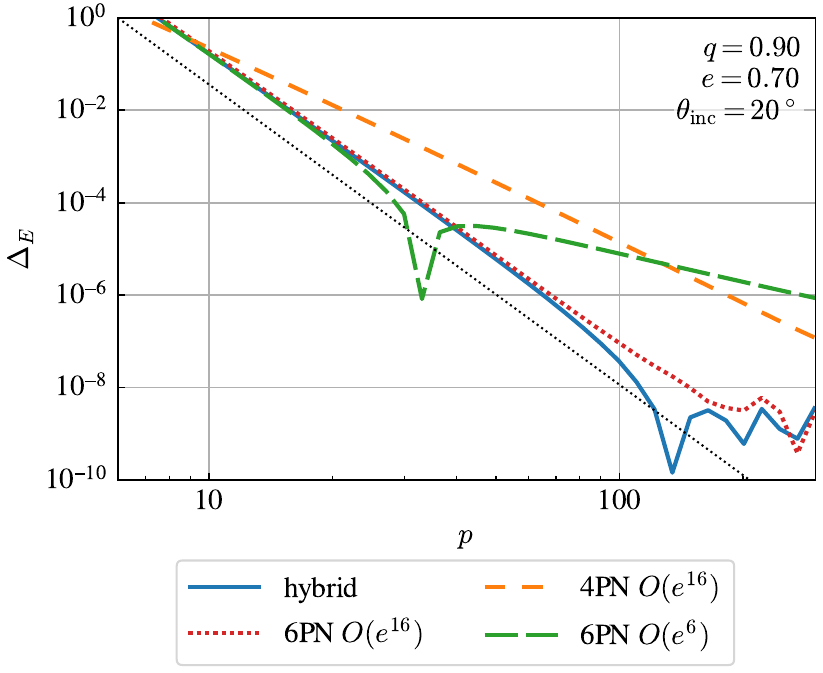}
\caption{
The relative errors in the hybrid analytic formulas of $\langle dE/dt \rangle_t$
as a function of the semi-latus rectum $p$ for $q=0.90$, $e=0.70$,
and $\theta_\mathrm{inc}=20^\circ$. 
The thin dotted line in each plot is reference proportional to $1/p^{13/2}=v^{13}$.
The hybrid formula (blue solid line) is constructed from the 4PN $O(e^{16})$ (orange short-dashed)
and 6PN $O(e^6)$ (green long-dashed) formulas (see Eq.~\eqref{eq:Edot_hybrid}). 
For comparison, the relative error of the 6PN $O(e^{16})$ formula is also shown (red dotted).}
\label{fig:DelE_hybrid_q09_e07_inc20}
\end{figure}

\begin{figure}[!ht]
\includegraphics[bb=0 0 841 576, width=0.95\linewidth]{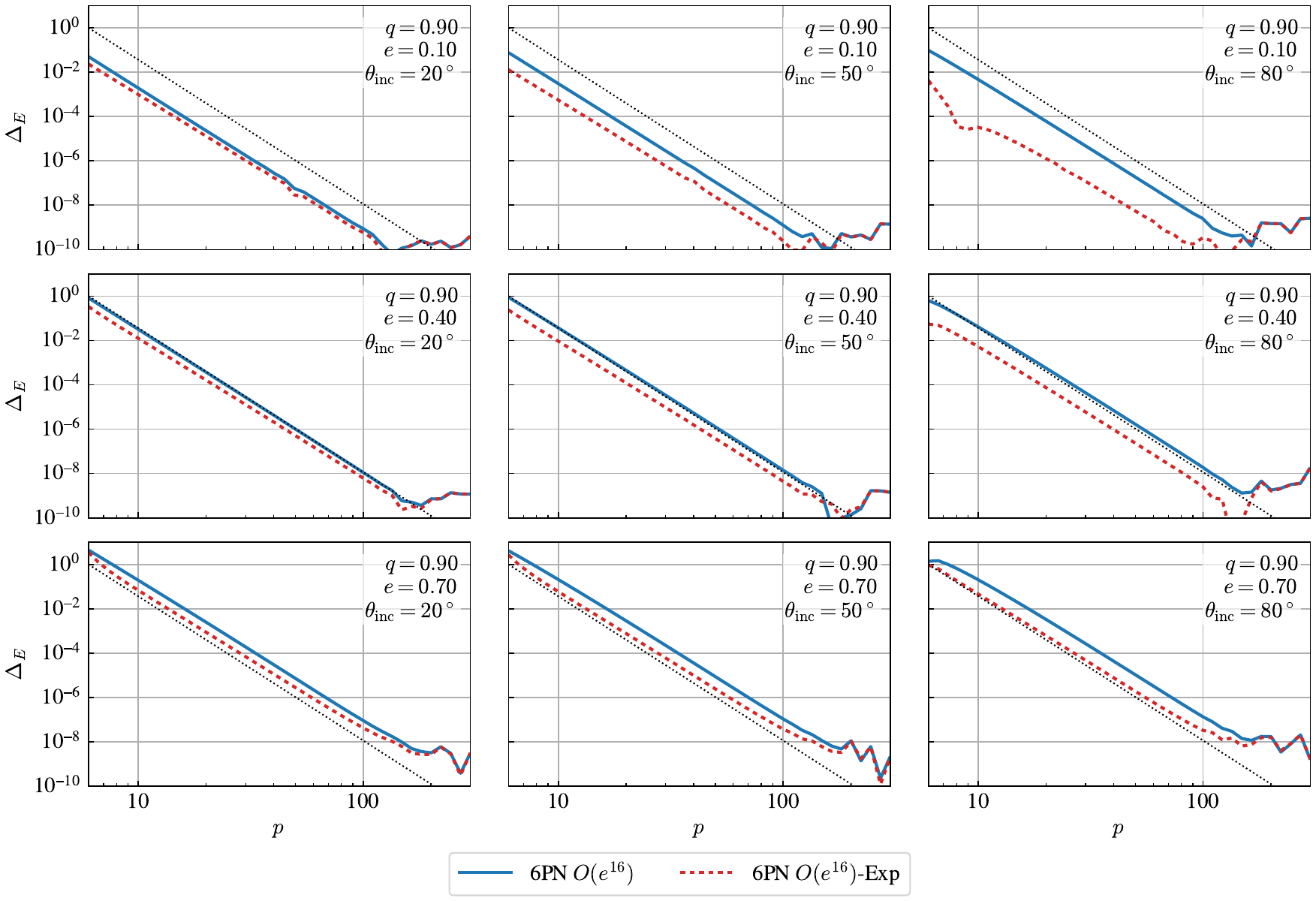}
\caption{Comparison of the PN formula and the exponential resummation formula given by Eq.~\eqref{eq:Idot_exp} 
for the secular change of the particle's energy $E$. 
In each panel, the relative errors in the 6PN $O(e^{16})$
formula (blue-solid line) and the exponential resummation formula (red-dotted line) are shown
as functions of $p$.
The cases with the eccentricity $e=0.10$, $0.40$, and $0.70$ are shown from the top to the bottom panels,
and those with the inclination angle $\theta_\mathrm{inc}=20^\circ$, $50^\circ$, and $80^\circ$
are shown from the left to the right panels.
We fix $q=0.90$ in all plots.}
\label{fig:DelE_resum_q09}
\end{figure}

\section{Attempts to improve accuracy of analytic formulas}
\label{sec:attempts}

\subsection{Hybrid formula}
\label{sec:hybrid}

Increasing both the PN order and the expansion order of $e$ simultaneously will require
enormous computational resources. To reduce the costs, as suggested in the previous
section, one may adopt a hybrid model combining the lower PN with higher order
of $e$-expansion and the higher PN with lower order of $e$
(to adjust the order of the PN expansion at each order of $e$-expansion).
As a demonstration, we construct a hybrid model with the 4PN $O(e^{16})$ and 6PN $O(e^6)$
formulas, which is expressed in the form of Eq.~\eqref{eq:Edot_e_exp} as
\begin{equation}
\left\langle \frac{dE}{dt} \right\rangle_t^\mathrm{Hybrid} :=
\left( \frac{dE}{dt} \right)_\mathrm{N}
\left[ \sum_{j=0}^{3} A_{2j}^{(12)} e^{2j}
+ \sum_{j=4}^{8} A_{2j}^{(8)} e^{2j}
\right] \,,
\label{eq:Edot_hybrid}
\end{equation}
where $A_{2j}$ are given up to the 6PN order (as $A_{2j}^{(12)}$) for $j=0,\,1,\,2,\,3$ and up to the 4PN
order (as $A_{2j}^{(8)}$) for $j=4,\,5,\,6,\,7,\,8$.
The relative errors of the hybrid formula comparing with the numerical results are
shown in Fig.~\ref{fig:DelE_hybrid_q09_e07_inc20}.
One can see that the hybrid formula (blue solid line in the plot) shows comparable
accuracy to the 6PN $O(e^{16})$ formula (red dotted line).
This suggests the potentiality of hybridization to reduce the computational costs.

\subsection{Exponential resummation}

One possible way to improve the accuracy of the PN formulas is to apply some resummation
methods, for example, Pad\'{e} approximation. In this work, we try to improve the accuracy
of the 6PN $O(e^{16})$ formulas by using the exponential resummation proposed in Ref.~\cite{Isoyama:2012bx}:
\begin{equation}
\left\langle \frac{dI}{dt} \right\rangle_t^\mathrm{res} :=
\left( \frac{dI}{dt} \right)_\mathrm{N} \exp{F_j^I} \,,
\label{eq:Idot_exp}
\end{equation}
where
\begin{equation}
F_j^I := \ln \left[ \left\langle \frac{dI}{dt} \right\rangle_t \Bigg/
\left( \frac{dI}{dt} \right)_\mathrm{N} \right]
\Bigg|_{\textrm{truncated after }j\textrm{-th order of }v} \,.
\label{eq:exponent_expresum}
\end{equation}
It should be noted that we do not perform the expansion with respect to $e$ in 
Eq.~\eqref{eq:exponent_expresum} while we do the expansion with respect to $v$.
Fig.~\ref{fig:DelE_resum_q09} shows the relative errors in the exponential resummation of the
6PN $O(e^{16})$ formula for $\langle dE/dt \rangle_t$.
The accuracy is not improved so much for most of cases shown in Fig.~\ref{fig:DelE_resum_q09}.
This result contrasts with the case of the 4PN $O(e^6)$ shown in our previous work~\cite{Sago:2015rpa},
in which the exponential resummation improves the accuracy by one order of magnitude.
The original idea of the exponential resummation \cite{Isoyama:2012bx} is to guarantee that the
energy flux is positive definite. The improvement of the accuracy is just a byproduct.
It suggests that the exponential resummation does not necessarily work well for all cases
although Ref.~\cite{Fujita:2014eta} has demonstrated the effectiveness for the 11PN calculation in the case of circular orbits.
Ref.~\cite{Munna:2020iju} has found that the exponential resummation (referred to as the logarithmic resummation in the paper)
begins to fail at relatively low eccentricity for the case of eccentric orbits in Schwarzschild spacetime.
To improve the accuracy of the current analytic formulas through any resummation techniques, 
we need further investigation.

\section{Summary}
\label{sec:summary}

In this paper, we analytically derived the secular evolution of the orbital parameters for a generic bound orbit in Kerr spacetime. 
Specifically, we computed the orbit-averaged rates of change for the energy, angular momentum, and the Carter constant. Operating within the framework of linear BHPT, we extended these analytical fluxes to the 6PN order and the 16th power of the orbital eccentricity at linear order in the mass ratio. Comparisons with numerical Teukolsky data confirm the validity of our formulas, with the relative error exhibiting the expected $p^{-13/2}$ scaling in the weak-field regime. 
While the 6PN ${O}(e^{16})$ expressions significantly improve accuracy at large orbital separations compared to previous 4PN and 5PN results, they do not monotonically improve accuracy in the strong-field (small-$p$) regime. This behaviour explicitly maps the known asymptotic nature of the PN series~\cite{Yunes:2008tw,Sago:2016xsp,Fujita:2017wjq,Munna:2020iju,Castillo:2024isq}, demonstrating that naive truncation at higher PN orders is insufficient to capture highly eccentric, strong-field dynamics.

To address this fundamental limitation and mitigate the computational bottlenecks of PN-GSF calculations, we proposed and validated a hybrid approximation. By strategically combining lower-order PN formulas with higher-order eccentricity corrections, and higher-order PN formulas with lower-order eccentricity corrections, this approach maintains high accuracy while drastically reducing the algebraic complexity, memory and CPU consumption required for the rapid evaluation of intermediate results. Additionally, we investigated the exponential resummation method~\cite{Isoyama:2012bx}; unlike its success at the 4PN order~\cite{Sago:2015rpa}, it yielded limited improvements at 6PN, indicating that alternative resummation techniques or direct hybridization with numerical GSF data may be necessary for higher-order fluxes.

To facilitate their application in broader waveform modelling efforts, the full analytical expressions are made publicly available at the Black Hole Perturbation Club (BHPC)~\cite{BHPC} for integration into community-driven resources like the Black Hole Perturbation Toolkit~\cite{BHPT,wardell_2025_15969633} and the LISA Waveform Working Group~\cite{LISAConsortiumWaveformWorkingGroup:2023arg}. 
In the near term, incorporating the proposed hybrid flux model into existing (post) adiabatic frameworks~\cite{Pound:2021qin,Fujita:2020zxe,Hughes:2021exa,Isoyama:2021jjd,Mathews:2025txc,Lewis:2025ydo} will enable the rapid generation of long-term orbital trajectories and phase evolution at the 6PN level. This provides a baseline for EMRI data analysis pipelines, where rapid coverage of the generic parameter space is crucial for exploring the non-local parameter degeneracies inherent to the EMRI signal space~\cite{Chua:2021aah,Chua:2022ssg}.

Looking ahead, several theoretical challenges remain to capture the strong-field dynamics more. 
Because our current formulation is already exact in both the primary BH spin and the orbital inclination, the natural next step is to relax the small-$e$ expansion. Deriving analytical PN formulas valid for arbitrary eccentricities---an effort currently underway at lower PN orders---will eliminate a major truncation error and significantly extend the validity of these models. 

Furthermore, our current formulation relies on standard multiscale analysis with phase-space averaging, which becomes incomplete when generic Kerr orbits encounter transient resonances between the radial and polar frequencies~\cite{Mino:2005an,Tanaka:2005ue,Flanagan:2010cd,vandeMeent:2013sza,Brink:2015roa,Lukes-Gerakopoulos:2021ybx} because oscillatory GSF effects fail to average out over the orbital timescale~\cite{Flanagan:2012kg,Isoyama:2018sib,Isoyama:2021jjd,Gupta:2022fbe,Lynch:2024ohd}. Extending the current PN-GSF treatments to account for resonant dynamics is an important next step, as LISA-relevant EMRIs will pass through at least one dynamically significant low-order resonance as they spiral through LISA's sensitive band~\cite{Ruangsri:2013hra}. 

Addressing these remaining challenges will maximize the inherent advantage of analytical formulations: the ability to evaluate orbital dynamics rapidly and continuously across a high-dimensional parameter space. By serving as a scalable complement to strong-field numerical data, we expect that these analytical frameworks provide a necessary foundation for high-precision GW astronomy.

\section*{Acknowledgment}

We are grateful to Alvin~J.~K.~Chua, Christian E.~A.~Chapman-Bird, Jezreel C.~Castillo, Chris Kavanagh, Josh Mathews and David Trestini for helpful discussions. 
This work was supported by JSPS KAKENHI Grant Numbers
JP21H01082, JP23K20845 (N.S., R.F., and H.N.), JP21K03582, and JP23K03432 (H.N.).
S.I. is supported by the Ministry of Education, Singapore, under the Academic Research Fund Tier 1 A-8001492-00-00 (FY2023).
%


\bibliographystyle{ptephy}
\bibliography{6PNe16}

\end{document}